\documentclass[%
aps,
prl,
 amsmath,amssymb,
reprint,
]{revtex4-1}

\usepackage{graphicx}% Include figure files
\usepackage{dcolumn}% Align table columns on decimal point
\usepackage{bm}% bold math
\usepackage{color}
\usepackage{amsmath}	% required for `\align' (yatex added)
\begin{document}

\newcommand{\diff}[2]{\frac{d#1}{d#2}}
\newcommand{\pdiff}[2]{\frac{\partial #1}{\partial #2}}
\newcommand{\fdiff}[2]{\frac{\delta #1}{\delta #2}}
\newcommand{\bx}{\bm{x}}
\newcommand{\bv}{\bm{v}}
\newcommand{\br}{\bm{r}}
\newcommand{\ba}{\bm{a}}
\newcommand{\by}{\bm{y}}
\newcommand{\bY}{\bm{Y}}
\newcommand{\bF}{\bm{F}}
\newcommand{\bn}{\bm{n}}
\newcommand{\be}{\bm{e}}
\newcommand{\new}{\nonumber\\}
\newcommand{\abs}[1]{\left|#1\right|}
\newcommand{\tr}{{\rm Tr}}
\newcommand{\HH}{{\mathcal H}}
\newcommand{\OO}{{\mathcal O}}
\newcommand{\var}{{\rm Var}}
\newcommand{\ave}[1]{\left\langle #1 \right\rangle}

\preprint{AIP/123-QED}

\title[Scaling theory of continuous symmetry breaking under
advection ]{Scaling theory of continuous symmetry breaking under
advection}
%Force line breaks With \\
\author{Harukuni Ikeda}
 \email{harukuni.ikeda@gakushuin.ac.jp}
\affiliation{
Department of Physics, Gakushuin University, 1-5-1 Mejiro, Toshima-ku, Tokyo 171-8588, Japan}

% \author{Pierfrancesco Urbani}
%  %Lines break automatically or can be forced with \\
% \affiliation{Institut de physique th\'eorique, Universit\'e Paris
% Saclay, CNRS, CEA, F-91191 Gif-sur-Yvette, France}

\date{\today}% It is always \today, today,

% \begin{abstract}
% Valid PACS numbers may be entered using the \verb+\pacs{#1}+ command.
% \end{abstract}
% \pacs{
% 64.60.Ht, %dynamic critical phenomena
% 05.20.-y, %classical statistical mechanics
% 05.20.Dd %Kinetic theory
% }% PACS, the Physics and Astronomy
% \pacs{Valid PACS appear here}% PACS, the Physics and Astronomy
%                              % Classification Scheme.


\begin{abstract}
In this work, we discuss how the linear and non-linear advection terms
modify the scaling behavior of the continuous symmetry breaking and
stabilize the long-range order, even in $d=2$ far from equilibrium, by
means of simple scaling arguments. For an example of the liner
advection, we consider the $O(n)$ model in the steady shear. Our scaling
analysis reveals that the model can undergo the continuous symmetry
breaking even in $d=2$ and, moreover, predicts the upper critical
dimension $d_{\rm up}=2$. These results are fully consistent with a
recent numerical simulation of the $O(2)$ model, where the mean-field
critical exponents are observed even in $d=2$. For an example of the
non-linear advection, we consider the Toner-Tu hydrodynamic theory,
which was introduced to explain polar-ordered flocks, such as the Vicsek
model. Our simple scaling argument reproduces the previous results by
the dynamical renormalization theory. Furthermore, we discuss the
effects of the additional non-linear terms discovered by the recent
re-analysis of the hydrodynamic equation. Our scaling argument predicts
that the additional non-linear terms modify the scaling exponents and,
in particular, recover the isotropic scaling reported in a previous
numerical simulation of the Vicsek model. We discuss that the critical
exponents predicted by the naive scaling theory become exact in $d=2$ by
using a symmetry consideration and similar argument proposed by Toner
and Tu.
\end{abstract}

\maketitle

%\section{Introduction}
\paragraph{Introduction.---}One of the most famous no-go theorems in equilibrium statistical
mechanics is the Mermin-Wagner theorem~\cite{mermin1966}, which
prohibits continuous symmetry breaking in $d=2$ dimensions.
However, the Mermin-Wagner theorem does not hold far from equilibrium,
and indeed, there are several out-of-equilibrium systems showing 
continuous symmetry breaking even in
$d=2$~\cite{vicsek1995,xy1995,reichl2010,loos2022long,dadhichi2020,leonardo2023,ikeda2023cor,ikeda2023does,ikeda2023harmonic}. 
A famous prototypical numerical model that breaks the Mermin-Wagner
theorem is the so-called Vicsek model~\cite{vicsek1995}. The model
consists of $XY$ spins flaying with a constant speed along their
magnetic direction, which mimics the flocking behavior among living
things such as birds and
bacteria~\cite{vicsek1995,nishiguchi2017}. Interestingly, the numerical
simulation of the model in $d=2$ showed that the model undergoes
continuous symmetry breaking from the disordered phase, where the mean
velocity of the spins vanishes $\ave{\bv}=0$, to the ordered phase,
where $\ave{\bv}\neq 0$~\cite{vicsek1995}.

In 1995, Toner and Tu proposed a hydrodynamic theory to explain the
long-range order of the Vicsek model in $d=2$~\cite{toner1995}.  The
Toner-Tu theory in 1995 (TT95) successfully explained the existence of
the order phase of the Vicsek model in $d=2$. However, subsequent
re-analysis of the hydrodynamic theory by J.~Toner reported several
additional terms that were not considered in the original theory
TT95~\cite{toner1995}. These terms hinder the exact calculations of the
critical exponents for compressible fluid~\cite{toner2012}. Recently, an
extensive numerical simulation of the Vicsek model has been
performed. The numerical results of the critical exponents are indeed
inconsistent with those of TT95. In particular, the numerical results
suggest almost isotropic scaling behavior~\cite{mahault2019}, while TT95
predicts anisotropic scaling~\cite{toner1995}. One of the purposes of
this work is to reconcile this discrepancy by taking into account the
effects of the additional non-linear terms.

In the hydrodynamic theory TT95, the key ingredient to stabilize the
long-range order in $d=2$ is the non-linear advection
$(\bv\cdot\nabla)\bv$, which leads to the super-diffusion and
facilitates the relaxation of the Goldstone
mode~\cite{toner1995,minami2022origin}. Interestingly, a recent
numerical simulation of the $O(n)$ model with steady shear revealed that
the linear advection $(\bv\cdot\nabla)\bm{\phi}$, where $\bm{\phi}$
denotes the order parameter, can also stabilize the long-range order and
allows the continuous symmetry breaking in $d=2$~\cite{nakano2021}. The
numerical result also reported that the critical exponents of the model
agree with the mean-field theory (linear analysis) even in $d=2$. Our
other goal is to explain this surprising result.

To tackle the above two problems, we perform simple scaling arguments for
the continuous symmetry breaking with advection. Our theory for the
linear advection of the $O(n)$ model predicts that the upper critical
dimension below which the Gaussian fixed point gets unstable is $d_{\rm
up}=2$. This explains the mean-field behavior observed in the numerical
simulation in $d=2$~\cite{nakano2021}. For the non-linear advection,
our scaling analysis can reproduce the previous theoretical
results reported in TT95~\cite{toner1995}. Furthermore, the simplicity
of the scaling argument allows us to take into account the additional
non-linear terms, which have not been taken into account in the original
theory TT95~\cite{toner1995}. The new critical exponents obtained by
this work indeed support the isotropic scaling reported in the
previous numerical simulation of the Vicsek model~\cite{mahault2019}.
We also argue that the exponents predicted by the native scaling
argument actually become exact in $d=2$.

%\section{$O(n)$ model in shear}
\paragraph{$O(n)$ model in shear.---}
We first discuss the effects of the linear advection. We
consider the $O(n)$ model in the steady shear~\cite{nakano2021}:
\begin{align}
\pdiff{\phi_a}{t}+\bm{v}\cdot\nabla\phi_a
= \nabla^2\phi_a
- \fdiff{F[\bm{\phi}]}{\phi_a}
+\xi_a,
\label{201609_30Dec23}
\end{align}
where the $n$-component vector $\bm{\phi}=\{\phi_1,\cdots, \phi_n\}$
denotes the order parameter, 
\begin{align}
F[\bm{\phi}] =  \int d\bx \left[\frac{\varepsilon}{2}\left(\bm{\phi}\cdot\bm{\phi}\right)
+ \frac{g}{4}\left(\bm{\phi}\cdot\bm{\phi}\right)^2\right]\label{062606_1Jan24}
\end{align}
denotes the standard $\phi^4$ free-energy, 
and $\xi_a$ denotes the thermal nosie of zero mean
and variance:
\begin{align}
\ave{\xi_a(\bx,t)\xi_b(\bx',t')} = 2T\delta_{ab}\delta(\bx-\bx')\delta(t-t').\label{021726_31Dec23}
\end{align}
The advection term in Eq.~(\ref{201609_30Dec23}),
$\bv\cdot\nabla\phi_a$, is a linear function of the order parameter.
For the velocity field $\bv$, we consider the simple shear along the
$x_1$ direction:
\begin{align}
\bm{v} = \dot{\gamma}x_2 \bm{e}_1,
\label{024236_31Dec23}
\end{align}
where $\bm{e}_1$ denotes the unit vector along the $x_1$ axis.
Recently, an extensive numerical simulation of the model has been
performed~\cite{nakano2021}.  The numerical result showed that the model
undergoes the continuous symmetry breaking even in
$d=2$. Furthermore, the critical exponents of the transition agree with
the mean-field prediction. Here, we explain this result through a simple
scaling argument.

To investigate the large spatio-temporal behavior of the model, we
consider the following scaling
transformations~\cite{nishimori2011elements,onuki1979}~\footnote{One can
introduce the proportional coefficients for the advection and diffusion
terms, as in $c_{a}\bv\cdot\nabla\phi_a$ and $c_{d}\nabla^2\phi_a$, and
consider their scaling dimensions. However, in such a case, the fixed
point corresponds to $c_a=c_d=0$: the model loses the spatial
dependence. We are not interested in such a case. For the same reason,
we also omit the proportional coefficient of $\nabla^2\bv$ and
$(\bv\cdot\nabla)\bv$ in the Vicsek model.}:
\begin{align}
&x_1 \to b^{\zeta_1}x_1,\
x_2 \to b^{\zeta_2}x_2,\
x_i \to b x_i\ i=3,\cdots, d,\new 
&t\to b^{z}t,\
\phi_a\to b^{\chi}\phi_a,\
\varepsilon\to b^{y_{\varepsilon}}\varepsilon,\
g\to b^{y_g}g.\label{094508_19Jan24} 
\end{align}
For the noise $\xi_a$, Eq.~(\ref{021726_31Dec23}) implies $\xi_a\to
b^{-\frac{z+\zeta_1+\zeta_2+d-2}{2}}\xi_a$~\cite{burger1989}.  Then,
Eq.~(\ref{201609_30Dec23}) reduces to
\begin{align}
&b^{\chi-z}\pdiff{\phi_a}{t}+ b^{\chi+\zeta_2-\zeta_1}\dot{\gamma}x_2 \pdiff{\phi_a}{x_1} \new 
&=b^{\chi-2\zeta_1}\pdiff{^2\phi_a}{x_1^2}
+b^{\chi-2\zeta_2}\pdiff{^2\phi_a}{x_2^2}
+b^{\chi-2}\sum_{i=3}^d\pdiff{^2\phi_a}{x_i^2}\new
&-b^{\chi+y_{\varepsilon}}\varepsilon\phi_a
- b^{y_g+3\chi}g \abs{\bm{\phi}}^2\phi_a 
+b^{-\frac{z+\zeta_1+\zeta_2+d-2}{2}}\xi_a\label{193835_7Jan24}
\end{align}
We first discuss the model without shear $\dot{\gamma}=0$. In this case,
assuming that each term in Eq.~(\ref{193835_7Jan24}) has the same scaling dimension,
we get
\begin{align}
&\chi-z = \chi-2\zeta_1= \chi-2\zeta_2 = \chi-2\new
&=\chi+y_{\varepsilon} = y_g + 3\chi = -(z+\zeta_1+\zeta_2+d-2)/2,
\end{align}
leading to
\begin{align}
z = 2, \zeta_1 = \zeta_2 = 1, \chi = \frac{2-d}{2}, y_{\varepsilon}=-2,\ y_g = d-4.\label{094954_19Jan24}
\end{align}
To see the stability of the ordered phase, one can observe the
fluctuation of the order parameter $\ave{\delta\phi_a^2}\sim
b^{2\chi}$. For $d\leq 2$, $\chi\geq 0$, and thus, the fluctuation
diverges in the thermodynamic limit $b\to\infty$, which destroys the
long-range order~\cite{toner1995}, as expected from the Mermin-Wagner
theorem~\cite{mermin1966}. Eq.~(\ref{094508_19Jan24}) implies that the
coefficient of the non-linear term $g$ scales as $g'=b^{-y_g}g$ after
the scale transformation $\bx'\to
b^{-1}\bx$~\cite{nishimori2011elements}. When $y_g>0$, the non-linear
term is irrelevant, and the critical exponents of the Gaussian fixed
point Eq.~(\ref{094954_19Jan24}) become exact. The upper critical
dimension $d_{\rm up}$ above which the Gaussian fixed point stabilizes
is obtained by setting $y_g=0$, leading to $d_{\rm
up}=4$~\cite{nishimori2011elements}. The results are fully consistent
with the scaling analysis of the equilibrium
model~\cite{nishimori2011elements}.

Now, we discuss how the finite shear rate $\dot{\gamma}\neq 0$ changes
the above scaling behavior. If we require all terms in
Eq.~(\ref{193835_7Jan24}) to have the same scaling dimension, we obtain
seven equations, while there are only six unknown variables, $\zeta_1$,
$\zeta_2$, $z$, $\chi$, $y_\varepsilon$, and $y_g$, which are
overdetermined and do not have solutions.  So, we here assume that the
advection ${\dot{\gamma}x_2\partial_{x_1}\phi_a\sim
b^{\chi+\zeta_2-\zeta_1}}$ plays a more dominant role than the diffusion
${\partial_{x_1}^2\phi_a\sim b^{\chi-2\zeta_1}}$, which is tantamount to
assuming ${\chi+\zeta_2-\zeta_1>\chi-2\zeta_1\to \zeta_1+\zeta_2>0}$ and
neglect the term $b^{\chi-2\zeta_1}$ in
Eq.~(\ref{193835_7Jan24}). Requiring that all the remaining terms in
Eq.~(\ref{193835_7Jan24}) have the same scaling dimension, we get
\begin{align}
&\chi-z = \chi+\zeta_2-\zeta_1 = \chi-2\zeta_2 = \chi-2 \new
&=
\chi+y_{\varepsilon} = y_g + 3\chi = -(z+\zeta_1+\zeta_2+d-2)/2,
\end{align}
leading to
\begin{align}
&z = 2, \zeta_1 = 3, \zeta_2 = 1, \chi = -\frac{d}{2},
y_{\varepsilon} = -2,\ y_g = d-2.\label{095348_9Jan24}
\end{align}
The assumption $\zeta_1+\zeta_2>0$ is satisfied self-consistently. The
anisotropic exponent $\zeta_1=3$ is consistent with the linear
analysis~\cite{nakano2021}. Since $\chi<0$, the fluctuation of the order
parameter $\ave{\delta\phi_a^2} \sim b^{2\chi}$ vanishes in the large
scale $b\gg 1$, which allows the long-range order even in $d=2$. This is
in sharp contrast with the equilibrium system without shear
$\dot{\gamma}=0$, where the long-range order can exist only for
$d>2$~\cite{mermin1966}. The upper critical dimension is obtained by
setting $y_g=0$~\cite{nishimori2011elements}, leading to $d_{\rm up} =
2$, meaning that Eqs.~(\ref{095348_9Jan24}) are exact even in $d=2$.
The exponents $\zeta_1$, $\zeta_2$, and $y_\varepsilon$ are indeed
consistent with a recent numerical simulation in
$d=2$~\cite{nakano2021}. However, strictly speaking, the logarithmic
corrections are expected at the upper critical dimension $d=d_{\rm
up}=2$~\cite{nishimori2011elements}. Further numerical studies would be
beneficial to elucidate this point.

%\section{Hydrodynamic theory for the Vicsek model}
\paragraph{Hydrodynamic theory for the Vicsek model.---}
Now we discuss the effects of the non-linear advection of the
hydrodynamic theory of the Vicsek model~\cite{vicsek1995}. We first
investigate the \textit{minimal} hydrodynamic theory for the polar
ordered flocks investigated in Ref.~\cite{basse2022}:
\begin{align}
\pdiff{\bm{v}}{t}+(\bm{v}\cdot\nabla) \bm{v}
= \nabla^2 \bm{v} -
\fdiff{F[\bm{v}]}{\bm{v}}
+\bm{\xi},\label{203525_31Dec23}
\end{align}
where $\bv(\bx,t)$ represents the local velocity of the flocks, and the
functional form of the free-energy $F$ is given by
Eq.~(\ref{062606_1Jan24}). Compared to the original model proposed by
Toner and Tu~\cite{toner1995}, the model Eq.~(\ref{203525_31Dec23})
assumes that the fluid is \textit{incompressible} and neglects the terms
related to the pressure-gradient forces, such as $\nabla p$ and
$\bv(\bv\cdot\nabla)p$. Some irrelevant terms, such as
$(\bv\cdot\nabla)^2\bv$, are also neglected~\cite{basse2022}. Since the
velocity field $\bv$ itself plays the role of the order parameter,
Eq.~(\ref{203525_31Dec23}) has the non-linear advection term
$(\bv\cdot\nabla)\bv$~\footnote{One can include two more non-linear
advection terms: $(\nabla\cdot\bv)\bv$ and $\nabla(\bv\cdot\bv)$, which
has the same scaling dimension as $(\bv\cdot\nabla)\bv$ and do not
change the results of the scaling analysis.}. In the ordered phase
$\varepsilon<0$, the mean value of the velocity field has a finite value
$\ave{\bv}=v_0\bm{e}_{\parallel}$, where $\bm{e}_{\parallel}$ denotes
the unit vector along the mean velocity $v_0=\abs{\ave{\bv}}$. Now we
decompose the velocity field as ${\bv=\bv_{\parallel}+ \bv_{\perp}}$. We
consider the following scaling transformations~\cite{toner1995}:
\begin{align}
&x_{\parallel} \to b^{\zeta}x_\parallel,\ \bx_{\perp} \to b\bx_{\perp},\
t\to b^{z}t,\new 
&\bv_{\perp}\to b^{\chi}\bv_{\perp}, \bv_{\parallel}\to b^{\chi'}\bv_{\parallel},
\varepsilon\to b^{y_{\varepsilon}}\varepsilon,\
g\to b^{y_g}g. 
\label{221607_31Dec23}
\end{align}
Then, Eq.~(\ref{203525_31Dec23}) in the space perpendicular to
$\bm{e}_{\parallel}$ reduces to
\begin{align}
&b^{\chi-z}\pdiff{\bv_{\perp}}{t}
+b^{2\chi-1}(\bv_{\perp}\cdot\nabla_{\perp})\bv_{\perp}
+b^{\chi+\chi'-\zeta}(\bv_{\parallel}\cdot\nabla_{\parallel})\bv_{\perp} \new 
&=
b^{\chi-2}\nabla_{\perp}^2 \bv_{\perp} 
+b^{\chi-2\zeta}\nabla_{\parallel}^2 \bv_{\perp} \new 
&-b^{\chi+y_{\varepsilon}}\varepsilon\bv_{\perp} 
-b^{y_g+\chi+2\chi'}g \abs{\bv_{\parallel}}^2\bv_{\perp}\new
&-b^{y_g+3\chi}g \abs{\bv_{\perp}}^2\bv_{\perp}
+b^{-\frac{z+\zeta+d-1}{2}}\bm{\xi}_{\perp}.\label{221947_31Dec23}
\end{align}
As before, we assume that the advection
${(\bv_\perp\cdot\nabla_\perp)\bv_\perp \sim b^{2\chi-1}}$ plays the
dominant role than the diffusion ${\nabla_\perp^2\bv_\perp\sim
b^{\chi-2}}$, which is tantamount to assuming ${2\chi-1>\chi-2\to
\chi>-1}$. Then, we obtain the following scaling relations:
\begin{align}
&\chi-z = \chi+\chi'-\zeta= 2\chi-1 = \chi-2\zeta \new 
&=y_\varepsilon+\chi= y_g+3\chi= -(z+\zeta+d-1)/2,
\end{align}
leading to
\begin{align}
&z = \frac{2(1+d)}{5},\ \zeta = \frac{d+1}{5},\ \chi = \frac{3-2d}{5},\new 
&
\chi' = -\frac{d+1}{5},\ 
 y_{\varepsilon} = -\frac{2(1+d)}{5},\ 
\label{235759_31Dec23} 
\end{align}
and
\begin{align}
y_{g} = \chi-z -\max[\chi+2\chi',3\chi] = \frac{2(d-4)}{5}.
\end{align}
The assumption $\chi>-1$ is satisfied self-consistently for $d<4$. The
resultant $z$, $\zeta$, and $\chi$ are consistent with the previous
study by Toner and Tu (TT95)~\cite{toner1995}.  The exponents $\chi$ and
$\chi'$ both become negative, $\chi<0$ and $\chi'<0$, for $d>3/2$, which
allows the long-range order even in $d=2$~\cite{toner1995}. Note that
they satisfy the scaling relation $\chi'=\chi-1+\zeta$, which is
expected from the condition of the incompressible flow $\nabla\cdot\bv =
\nabla_{\parallel}\cdot\bv_\parallel + \nabla_{\perp}\cdot\bv_\perp
=0$~\cite{chen2016mapping}. By setting $y_g=0$, we get the upper
critical dimension $d_{\rm up}=4$. The previous renormalization group
analysis revealed that there are no corrections for the exponents $z$,
$\zeta$, and $\chi$ for incompressible
fluid~\cite{toner1995,toner2005,chen2018}.

\begin{table}[t]
\begin{center}
\caption{Critical exponents. Data are taken from
Ref~\cite{mahault2019}.}  \label{235502_31Dec23}
\begin{tabular}{l|rrr|rrr}
&$d = 2$ &  & & $d=3$ & & \\ 
& TT95&Vicsek& This work & TT95&Vicsek& This work \\ \hline
$\chi$ & -1/5 & -0.31(2)& -1/3& -3/5 & -0.62 & -2/3\\
$\zeta$ & 3/5 & 0.95(2) &1 & 4/5 & 1 &1 \\
$z$ & 6/5& 1.33(2)& 4/3 & 8/5& 1.77 & 5/3 
\end{tabular}
\end{center}
\end{table}

Recently, an extensive numerical simulation of the Vicsek model reported
that the critical exponents $\chi$, $\zeta$, and $z$ are inconsistent
with TT95, Eqs.~(\ref{235759_31Dec23}), see
TABLE~\ref{235502_31Dec23}~\cite{mahault2019,nishiguchi2023}. In
particular, the numerical results suggest the almost isotropic scaling
$\zeta\approx 1$~\cite{mahault2019,nishiguchi2023}. This discrepancy
would be reconciled by considering the hydrodynamic theory for
\textit{compressible} fluid~\cite{toner2012}. For the compressible
fluid, the density fluctuations $\delta\rho$ are coupled with $\bv$,
leading to some additional non-linear coupling terms between $\bv$ and
$\delta\rho$ in Eq.~(\ref{221947_31Dec23})~\cite{toner1995,toner2012}.
These non-linear terms were not taken into account in the previous theory
(TT95)~\cite{toner1995,toner2012}.

For the scaling argument, it is sufficient to estimate the order of
magnitude of the new non-linear terms for $b\gg 1$. Below, we provide a
rough estimation. In principle, one can express $\delta\rho$ as a
function of $\bv$ by solving the equation of continuity $\partial_t \rho
= -\nabla\cdot \bm{J}$~\cite{toner1995,toner2012}, meaning that the
additional non-linear terms would also be expressed as functions of
$\bv$. The effects of the density fluctuations appear through the
pressure-gradient forces, such as $\nabla p$ and $\bv(\bv\cdot\nabla
p)$~\cite{toner2012}, implying that the new non-linear terms may also
involve at least one spatial derivative $\nabla_\parallel$ or
$\nabla_\perp$. So, the least-order non-linear terms would be written as
\begin{align}
(\nabla\cdot\bv)\bv_{\parallel{\rm or}\perp},\ 
(\bv\cdot\nabla)\bv_{\parallel{\rm or}\perp},\
\nabla_{\parallel{\rm or}\perp}(\bv\cdot\bv).
 \label{154210_17Jan24}
\end{align}
The higher order terms of $\bv_{\parallel{\rm or}\perp}$ and
$\nabla_{\parallel{\rm or}\perp}$ are negligible since these terms have
smaller scaling dimensions.  From the scaling point of view,
Eqs.~(\ref{154210_17Jan24}) yield contributions proportional to
$b^{2\chi-1}$, $b^{\chi+\chi'-1}$, $b^{2\chi'-1}$, $b^{2\chi-\zeta}$,
$b^{\chi+\chi'-\zeta}$, and $b^{2\chi'-\zeta}$.  Using
Eqs.~(\ref{235759_31Dec23}), one can see that most terms are
irrelevant or do not change the scaling. However,
\begin{align}
\nabla_{\parallel}(\bv_\perp\cdot\bv_\perp)
\sim b^{2\chi-\zeta}
\label{231541_15Jan24}
\end{align}
diverge much faster than the terms in Eq.~(\ref{221947_31Dec23}) for
$b\gg 1$, which makes the scaling Eqs.~(\ref{235759_31Dec23}) improper.
The above discussion showed that the most relevant contribution is
$b^{2\chi-\zeta}$. Note that the argument itself does not guarantee the
existence of such terms. However, fortunately, the non-linear terms having
the same scaling dimension have indeed been found by a more detailed
analysis of the hydrodynamic theory of the compressible fluid by
J.~Toner~\cite{toner2012}. He reported several non-linear terms coupled
to the density fluctuations $\delta \rho$, such as
$\delta\rho\partial_\parallel \bv_\perp$ and
$\bv_\perp\partial_\parallel\delta\rho$, where $\delta\rho$ and
$\bv_\perp$ have the same scaling dimension $\delta\rho \sim \bv_\perp
\sim b^{\chi}$, implying $\delta\rho\partial_\parallel \bv_\perp
\sim\bv_\perp\partial_\parallel\delta\rho\sim
b^{2\chi-\zeta}$~\cite{toner2012}.

% Interestingly, the numerical result suggestthe scaling
% relation also hold even in $d=3$, suggesting that the relevant non-linear
% terms can also be written as total derivative of $\parallel$ or $\perp$
% in $d=3$~\cite{mahault2019}.

Now we discuss the additional terms indeed recover the isotropic scaling
$\zeta=1$. To get the new scaling exponents, we assume that the
advection-like terms along the parallel direction,
Eqs.~(\ref{231541_15Jan24}), play more dominant roles than the diffusion
along that direction $\nabla_\parallel^2 \bv_\perp\sim
b^{\chi-2\zeta}$, which is tantamount to assuming $2\chi-\zeta>\chi-2\zeta
\to \chi+\zeta >0$. Then, we get the new scaling relations: 
\begin{align}
&\chi-z =
\chi+\chi'-\zeta = 2\chi-1 = 2\chi-\zeta\new 
&=y_\varepsilon+\chi = y_g +
3\chi = -(z+\zeta+d-1)/2,
\end{align}
leading to the isotropic scaling
exponents
\begin{align}
&z = \frac{d+2}{3},\ \zeta = 1,\ \chi = \chi'= \frac{1-d}{3},\new
&y_\varepsilon = -\frac{2+d}{3},\ y_g = \frac{d-4}{3}.\label{235741_31Dec23} 
\end{align}
The assumption $\chi+\zeta>0$ is satisfied self-consistently for $d<4$.
By setting $y_g=0$, we get the upper critical dimension $d_{\rm up}=4$.
For $d<4$, our theory may not give the correct result near the
transition point. Furthermore, the transition of the Vicsek model is
known to be discontinuous in $d=2$, implying that one can not apply the
scaling argument itself~\cite{gregoire2004,solon2015}. Deep inside the
ordered phase, the fluctuations are predominantly controlled by the
Goldstone modes, which do not change the free-energy
$F[\bv]$. Therefore, the non-linear term proportional to $g$ would be
negligible. In TABLE~\ref{235502_31Dec23}, we compare the
Eqs.~(\ref{235741_31Dec23}) with the recent numerical results of the
Vicsek model in the order phase in $d=2$ and $3$. The agreement is
reasonably good, considering the simplicity of our theory. Further
theoretical~\cite{toner1995,toner2005,chen2016mapping,sartori2019,tasaki2020,diCarlo2022,jentsch2023},
numerical~\cite{gregoire2004,mahault2019}, and, hopefully,
experimental~\cite{nishiguchi2017,iwasawa2021,nishiguchi2023} studies
would be beneficial to estimate more accurate values of the critical
exponents.

\paragraph{Exact critical exponents in $d=2$.---}
We get almost perfect agreement in $d=2$. This is surprising because (i)
the full hydrodynamic equation involves far more non-linear terms than
those considered in the previous section~\cite{toner2012}, and (ii) the
naive scaling argument, in general, does not give the exact critical
exponents for non-linear systems~\cite{nishimori2011elements}.  Below, we
try to justify this result by using symmetry
consideration~\cite{sartori2019} and argument in
TT95~\cite{toner1995,toner2005}.  First, note that the fluctuations in
the ordered phase are dominated by the Goldstone mode $\bv_\perp$, which
does not change the free-energy $F[\bv]$. As before, we assume that the
density fluctuation $\delta\rho$ is expressed as a function of
$\bv\approx \bv_\perp$. Then, $\bv_\perp$ is only the relevant quantity.
Let us assume that the equation of motion (EOM) of $\bv_\perp$ is written as
follows:
\begin{align}
\pdiff{\bv_\perp}{t} = \bm{f}[\bv_\perp] + \bm{\xi}_\perp,\label{055937_20Jan24}
\end{align}
where $\bm{f}$ denotes the restoring force of $\bv_\perp$, and
$\bm{\xi}_\perp$ denotes the white noise.  The restoring force of the
Goldstone mode $\bm{f}$ should vanish in the limit of the
small wave number in the Fourier space, implying that each term in
$\bm{f}$ should involve at least one spatial
derivative $\nabla_{\perp{\rm or}\parallel}$.  Above the lower critical
dimension, $\bv_\perp\sim b^{\chi}$, $\nabla_\perp\sim b^{-1}$, and
$\nabla_\parallel\sim b^{-\zeta}$ vanish in the large scale $b\to\infty$. So,
$\bm{f}[\bv_\perp]$ would be expanded by $\bv_\perp$ and $\nabla_{\perp{\rm
or}\parallel}$ as follows~\cite{sartori2019}:
\begin{align}
\bm{f}[\bv_\perp] &= (D_1\nabla_\perp^2+D_2\nabla_\parallel^2)\bv_\perp
 +(D_3\nabla_\perp+D_4\nabla_\parallel)(\nabla_\perp\cdot\bv_\perp)\new 
 &+\lambda_1 (\bv_{\perp}\cdot\nabla_\perp)\bv_{\perp}
 +\lambda_2 (\nabla_\perp\cdot \bv_{\perp})\bv_{\perp}
+\lambda_3 \nabla_\perp (\bv_{\perp}\cdot\bv_{\perp})\new 
& +\nu\nabla_\parallel (\bv_{\perp}\cdot\bv_{\perp})+O(\abs{\nabla}^4,\abs{\bv_\perp}^4),\label{055602_21Jan24}
\end{align}
where $D_i$, $\lambda_i$, and $\nu$ are some constants. The higher order terms of
$\bv_\perp$ and $\nabla_{\perp{\rm or}\parallel}$ have smaller scaling
dimensions and would be negligible for $b\gg 1$. In $d=2$, $\bv_{\perp}$
has just one component, and the equation can be further simplified as
\begin{align}
\pdiff{v_\perp}{t} &=
 (d_1\partial_\perp^2 + d_2\partial_\parallel^2
+d_3\partial_\perp\partial_\parallel)v_\perp,\new 
 &+\lambda \partial_{\perp}v_\perp^2
 + \nu \partial_\parallel v_\perp^2  
 + \xi_\perp,\label{141706_18Jan24}
\end{align}
where $d_i$, $\lambda$, and $\nu$ are some constants. For $\nu=0$, the
critical exponents of Eq.~(\ref{141706_18Jan24}) have been calculated
exactly~\cite{sartori2019}, leading to the same results as TT95,
Eqs.~(\ref{235759_31Dec23}). Below, we derive the exact critical exponents
for $\nu\neq 0$. Since the non-linear terms proportional to $\lambda$ and
$\mu$ are written as the total derivative of $\parallel$ or $\perp$, the
perturbative expansion of these terms can only yield the terms 
$\partial_{\parallel{\rm or}\perp}({some\ function})$.  This implies
that the perturbative re-normalization calculation for these non-linear
terms does not contribute to the terms that do not involve
the spatial derivatives~\cite{toner1995,mahault2019,sartori2019}.  In
particular, the scaling behaviors of $\partial_{t}v_\perp\sim
b^{\zeta-z}$ and $\xi_\perp \sim b^{-(z+\zeta+d-1)/2}$ should remain
unchanged.  Assuming $\partial_t v_{\perp} \sim \xi_\perp$, we
get~\cite{mahault2019}:
\begin{align}
 z = 2\chi + \zeta + d-1.\label{142909_18Jan24}
\end{align}
A recent numerical simulation of the Vicsek model indeed confirmed the
above scaling relation in $d=2$~\cite{mahault2019}. One can see that
Eq.~(\ref{141706_18Jan24}) is invariant under the ``Pseudo-Galilean''
transformation: $v_\perp\to v_\perp + V$, $x_\perp \to x_\perp +
2\lambda Vt$, and $x_\parallel\to x_\parallel + 2\nu Vt$, which implies
the following scaling relations~\cite{toner1995,toner2005}
\begin{align}
&1= \chi + z,\
\zeta= \chi + z,\new 
&\to \zeta = 1,\ \chi = 1-z.
\label{142914_18Jan24} 
\end{align}
Using Eqs.~(\ref{142909_18Jan24}) and (\ref{142914_18Jan24}), we
reproduce the exponents $z$, $\zeta$, and $\chi$ in
Eqs.~(\ref{235741_31Dec23}). Since Eq.~(\ref{141706_18Jan24}) is the
most general form of EOM obtained by the symmetry consideration, we
conclude that $z$, $\zeta$, and $\chi$ calculated above are the exact
critical exponents of the Vicsek model in $d=2$. It is unclear if a
similar argument can be applied to Eq.~(\ref{055937_20Jan24}) in $d=3$.

%\section{Summary}
\paragraph{Summary.---}
In summary, we have investigated the continuous symmetry breaking with
linear and non-linear advections by means of simple scaling
arguments. Our theory demonstrated that the linear and non-linear
advection terms can generally reduce the lower-critical dimension and
stabilize the long-range order even in $d=2$, where the Mermin-Wanger
theorem prohibits the long-range order in
equilibrium~\cite{mermin1966}. In particular, our scaling theory, for
the first time, can explain the mean-field behavior of the sheared
$O(2)$ model in $d=2$~\cite{nakano2021} and the isotropic scaling
behavior of the Vicesek model~\cite{mahault2019}.

% Both scaling theories become exact in $d=2$. For the $O(n)$ model in the
% steady shear, this is because the advection term reduce the upper
% critical dimension to $d_{\rm up}=2$, meaning that the non-linear term
% become irrevant. However, strictly speaking, the logarithmic corrections
% are expected on the upper critical dimension $d=d_{\rm
% up}=2$~\cite{nishimori2011elements}. Further numerical studies would be
% beneficial to elucidate this point. For the hydrodynamic theory for
% polar-ordered flocks, the reason to be exact is more complicated: this
% is a consequence of the Pseudo-Galilean invariance and the fact that the
% non-linear terms are written as total derivatives. The agreement between
% the theory and numerical simulation of the Vicsek model is almost
% perfect in $d=2$, see Table~\ref{235502_31Dec23}, but there is a small
% deviation for the isotropic exponent $\zeta$. Further numerical and
% hopefully experimental studies would be beneficial to elucidate this
% point~\cite{iwasawa2021}.

\acknowledgments We thank H.~Nakano, Y.~Kuroda, and D.~Nishiguchi for
valuable discussions and comments. This work was supported by KAKENHI
23K13031.

% \subsection*{DATA AVAILABILITY} 
% The data that supports the findings of this study are available within the article.

% The data that support the findings of this study are available from the
% corresponding author upon reasonable request.

\bibliography{reference}

\end{document}